\documentclass[12pt]{article}

\pdfoutput=1

\usepackage{scicite}

\usepackage{times}

\usepackage[pdftex,final]{graphicx}     

\newenvironment{sciabstract}{%
\begin{quote} \bf}
{\end{quote}}

\newcommand\spinup{|\!\!\uparrow\rangle}
\newcommand\spindown{|\!\!\downarrow\rangle}
\newcommand\spinupdown{|\!\!\uparrow,\downarrow\rangle}
\newcommand\spindownup{|\!\!\downarrow,\uparrow\rangle}
\newcommand\spinpairleft{|\!\!\uparrow\downarrow,0\rangle}
\newcommand\spinpairright{|0,\uparrow\downarrow\rangle}
\renewcommand\vec[1]{\mathbf{#1}}

\newcounter{lastnote}
\newenvironment{scilastnote}{%
\setcounter{lastnote}{\value{enumiv}}%
\addtocounter{lastnote}{+1}%
\begin{list}%
{\arabic{lastnote}.}
{\setlength{\leftmargin}{.22in}}
{\setlength{\labelsep}{.5em}}}
{\end{list}}

\title{Time-resolved Observation and Control of Superexchange Interactions with Ultracold Atoms in Optical Lattices} 

\author
{S. Trotzky,$^{1\dagger}$ P. Cheinet,$^{1\dagger}$ S. F\"olling,$^{1}$ M. Feld,$^{1,2}$\\ U. Schnorrberger,$^{1}$ A. M. Rey,$^{3}$ A. Polkovnikov,$^{4}$ E. A. Demler,$^{3,5}$\\ M. D. Lukin,$^{3,5}$ and I. Bloch$^{1\ast}$\\
\\
\normalsize{$^{1}$Institut f\"ur Physik, Johannes Gutenberg-Universit\"{a}t, 55099 Mainz, Germany}\\
\normalsize{$^{2}$Fachbereich Physik, Technische Universit\"{a}t Kaiserslautern, 67663 Kaiserslautern, Germany}\\
\normalsize{$^{3}$Institute for Theoretical Atomic, Molecular and Optical Physics,}\\
\normalsize{Harvard-Smithsonian Center of Astrophysics, Cambridge, MA, 02138, USA}\\
\normalsize{$^{4}$Department of Physics, Boston University, Boston, MA, 02215, USA}\\
\normalsize{$^{5}$Physics Department, Harvard University, Cambridge, MA, 02138, USA} 
\\
\\
\normalsize{$^\ast$To whom correspondence should be addressed; E-mail:  bloch@uni-mainz.de.}\\
\normalsize{$^\dagger$These authors contributed equally to this work.}
}

\date{}

\begin{document} 

\baselineskip20pt

\maketitle

\begin{sciabstract} 
Quantum mechanical superexchange interactions form the basis of quantum magnetism in strongly correlated electronic media. We report on the direct measurement of superexchange interactions with ultracold atoms in optical lattices. After preparing a spin-mixture of ultracold atoms in an antiferromagnetically ordered state, we measure a coherent superexchange-mediated spin dynamics with coupling energies from 5\,Hz up to 1\,kHz. By dynamically modifying the potential bias between neighboring lattice sites, the magnitude and sign of the superexchange interaction can be controlled, thus allowing the system to be switched between antiferromagnetic or ferromagnetic spin interactions. We compare our findings to predictions of a two-site Bose-Hubbard model and find very good agreement, but are also able to identify corrections which can be explained by the inclusion of direct nearest-neighbor interactions.
\end{sciabstract}

Quantum spin systems on a lattice have served for decades as paradigms for condensed matter and statistical physics, elucidating fundamental properties of phase transitions and acting as models for the emergence of quantum magnetism in strongly correlated electronic media. In all these cases, the underlying systems rely on a spin-spin interaction between particles on neighboring lattice sites, such as in the Ising or Heisenberg model \cite{Ising:1925,Heisenberg:1926,Auerbachbook}. As initially proposed for electrons by Dirac \cite{Dirac:1926,Dirac:1929} and Heisenberg \cite{Heisenberg:1926,Heisenberg:1928}, effective spin-spin interactions can arise due to the interplay between the spin-independent Coulomb repulsion and exchange symmetry and do not require any direct coupling between the spins of the particles. The nature of such spin-exchange interactions is typically short-ranged, as it is governed by the wave function overlap of the underlying electronic orbitals. In several topical insulators, such as ionic solids like e.g. CuO and MnO, however, antiferromagnetic order arises even though the wave function overlap between the magnetic ions is practically zero. In this case a "superexchange" interaction mediated by higher order virtual hopping processes can be effective over large distance \cite{Kramers:1934,Anderson:1950} which leads to an (anti)-ferromagnetic coupling between bosons (fermions) on neighboring lattice sites \cite{Auerbachbook}. Such superexchange interactions are believed to play an important role in the context of high-$T_c$ superconductivity \cite{Lee:2006}. Furthermore, they can form the basis for the generation of robust quantum gates similar to recent work in electronic double quantum dot systems \cite{Burkard:1999,Petta:2005}, and can be used for the efficient generation of multi-particle entangled states \cite{Rey:2007,Briegel:2001}, as well as for the production of many-body quantum phases with topological order \cite{Kitaev:2006,Duan:2003,Santos:2004}.
 
We report on the direct observation of superexchange interactions with ultracold atoms in optical lattices \cite{Lewenstein:2007, Eckert:2007}. Previous experiments have shown that spin-spin interactions between neighboring atoms can be implemented in discrete time steps \cite{Jane:2003,Sorensen:1999} by bringing the atoms together on a single site and carrying out controlled collisions \cite{Jaksch:1999,Sorensen:1999,Mandel:2003b} or onsite exchange interactions \cite{Anderlini:2007}. The superexchange interactions demonstrated here, however, directly implement nearest-neighbor spin interactions in the many-body system and allow for a continuous "analog" simulation of spin lattice Hamiltonians.

We probe the superexchange interactions by first preparing two atomic spin states of $^{87}$Rb in an antiferromagnetic order \cite{Neel:1936} and then recording the time evolution of the spins of neighboring atoms in isolated double well potentials \cite{Sebby:2006,Sebby:2007a,Foelling:2007a} for weak to strong tunnel couplings. For dominating onsite interactions over the tunnel coupling between lattice sites, we find pronounced sinusoidal spin oscillations due to an effective Heisenberg-type superexchange Hamiltonian, whereas for stronger tunnel coupling a more complex dynamics emerges. In addition, we show how the strength and sign of the superexchange interaction can be directly controlled by introducing a potential bias between neighboring wells. Furthermore, we find that corrections to the two-site Bose-Hubbard model (BHM) which take into account the direct interaction between particles on neighboring lattice sites are needed to fully explain our data.

\paragraph*{Theoretical model.} 
An isolated system of two coupled potential wells constitutes the simplest concept for the investigation of superexchange-mediated spin dynamics between neighboring atoms. We consider a single double well potential occupied by a pair of bosonic atoms with two different spin-states denoted by $\spinup$ and $\spindown$. If the vibrational level splitting in each well is much larger than all other relevant energy scales and intersite interactions are neglected, the system can be described in a two-mode approximation by a two-site version of the Bose-Hubbard Hamiltonian
\begin{equation}
    \hat H
  = \sum_{\sigma = \uparrow,\downarrow}
          \left[ -J\left(\hat a^\dagger_{\sigma{\rm L}} \hat a_{\sigma{\rm R}}
                +\hat a^\dagger_{\sigma{\rm R}} \hat a_{\sigma{\rm L}}\right)
                -\frac{1}{2}\,\Delta\,(\hat n_{\sigma{\rm L}}-\hat n_{\sigma{\rm R}})\right]
    +U\left(\hat n_{\uparrow{\rm L}} \hat n_{\downarrow{\rm L}}
                + \hat n_{\uparrow{\rm R}} \hat n_{\downarrow{\rm R}}\right)\,,
\end{equation}
where the operators $\hat a^\dagger_{\sigma{\rm L,R}}$ and $\hat a_{\sigma{\rm L,R}}$ create and annihilate an atom with spin $\sigma$ in the left and right well respectively, $\hat n_{\sigma{\rm L,R}}$ count the number of atoms per spin state and well, $J$ is the tunnel matrix element, $\Delta$ the potential bias or tilt along the double well axis and $U=U_{\uparrow\downarrow}=g \times \int w^4_{\rm L,R}(\vec x) d^3 x$ the onsite interaction energy between two atoms in $\spinup$ and $\spindown$. Here, $g = (4\pi\hbar^2 a_{\rm s}^{\uparrow\downarrow})/M$ is the effective interaction strength with $a_s^{\uparrow\downarrow}$ being the positive scattering length for the spin states used in the experiment, $M$ is the mass of a single atom and $w_{\rm L,R}(\vec x)$ denote the wave functions for a particle localized on the left or right side of the double well. The state of the system can be described as a superposition of the Fock states $\{ \spinupdown, \spindownup, \spinpairleft, \spinpairright\}$, where the left and right side in the notation represent the occupation of the left and right well, respectively, and the states $\spinpairleft$ and $\spinpairright$ are spin triplet states. In the following, we will focus on the dynamical evolution of the population imbalance $x = n_{\rm L} - n_{\rm R}$ and the N\'{e}el order parameter or "spin imbalance" $N_z = (n_{\rm\uparrow L}+n_{\rm\downarrow R}-n_{\rm\uparrow R}-n_{\rm\downarrow L})/2$ starting with double wells initially prepared in $\spinupdown$. Here $n_{\rm\uparrow,\downarrow;L,R}=\langle \hat n_{\rm\uparrow,\downarrow;L,R} \rangle$ denote the corresponding quantum mechanical expectation values and $n_{\rm L,R}=n_{\rm \uparrow L,R}+n_{\rm\downarrow L,R}$.

In the limit of dominating interactions ($U \gg J$), when starting in the subspace of singly occupied wells spanned by $\spinupdown$ and $\spindownup$, the energetically high lying states $\spinpairleft$ and $\spinpairright$ can only be reached as "virtual" intermediate states in second order tunneling processes. Such processes lead to a non-local (super) spin-exchange interaction, which couples the states $\spinupdown$ and $\spindownup$ (see Fig.~1A). More generally, for an arbitrary spin configuration with equal interaction energies $U_{\uparrow\uparrow} = U_{\uparrow\downarrow} = U_{\downarrow\downarrow}$ (see ref.~\cite{ScattLengthNote}), the second order hopping events are described by an isotropic Heisenberg-type effective spin Hamiltonian in the limit $U \gg J$ \cite{Auerbachbook,Duan:2003,Kuklov:2003,Altman:2003}:
\begin{equation}
	  \hat H_{\rm eff} 
	= -2 J_{\rm ex} \hat S_{\rm L} \cdot \hat S_{\rm R} = 
	  -J_{\rm ex}\left( \hat S_{\rm L}^+ \hat S_{\rm R}^- + \hat S_{\rm L}^- \hat S_{\rm R}^+\right) -2J_{\rm ex} \hat S_{\rm L}^z \hat S_{\rm R}^z\,,
\end{equation}
where $\hat S_{\rm L,R}^+=\spinup\langle \downarrow \!\!|_{\rm L,R}$, $\hat S_{\rm L,R}^-=\spindown\langle \uparrow \!\!|_{\rm L,R}$ and $\hat S_{\rm L,R}^z = (\hat n_{\rm\uparrow L,R}-\hat n_{\rm\downarrow L,R})/2$ denote the corresponding spin operators of the system, with $\hat S_{\rm L,R}^\pm=\hat S_x \pm i \hat S_y$. The effective coupling strength $J_{\rm ex}$ represents the superexchange and can readily be evaluated by perturbation theory up to quadratic order in the tunneling operator which yields $J_{\rm ex}=2J^2/U$.

When a potential bias $\Delta > 0$ is applied, the degeneracy of the two intermediate states in the superexchange process is lifted (see Fig.~1A). For $J, \Delta \ll U$ this leads to a modification of the effective superexchange coupling with now $J_{\rm ex} = J^2/(U+\Delta) + J^2/(U-\Delta) = 2J^2U / (U^2-\Delta^2)$ \cite{Duan:2003}. By tuning the bias to $\Delta > U$, it is possible to change the sign of $J_{\rm ex}$ and therefore to switch between ferromagnetic and antiferromagnetic superexchange interactions. For $J \ll |U-\Delta|$ the picture of an effective coupling via two virtual intermediate states is again valid and the full reversal to $J_{\rm ex} = -2J^2/U$ is found to be reached for $\Delta = \sqrt{2}\,U$.

For symmetric double wells ($\Delta = 0$), the Hamiltonian Eq.~1 can be diagonalized analytically to give a valid picture for all values of $J$ and $U$ within the single band BHM. A convenient basis is given by the spin triplet and singlet state $|t/s\rangle = (\spinupdown\pm\spindownup)/\sqrt{2}$ and the states $|\pm\rangle \equiv (\spinpairleft\pm\spinpairright)/\sqrt{2}$. Two of the eigenstates are linear combinations of $|t\rangle$ and $|+\rangle$, where the one having the larger overlap with $|t\rangle$ is the ground state. The spin singlet $|s\rangle$ and the state $|-\rangle$ are already eigenstates themselves with energy $0$ and $U$ respectively (see Fig.~1B). As a direct consequence, $|-\rangle$ cannot be reached from the initial state $\spinupdown = (|s\rangle + |t\rangle)/\sqrt{2}$. Therefore, the dynamical evolution of the spin imbalance contains only two frequencies
\begin{equation}\label{eq:Frequencies}
  \hbar \omega_{1,2} = \frac{U}{2}\left(\sqrt{\left(\frac{4J}{U}\right)^2+1} \pm 1\right)\, .
\end{equation}
The extraction of these frequencies from time-resolved measurements allows for the determination of $2J=\hbar\sqrt{\omega_1\omega_2}$ and $U=\hbar(\omega_1-\omega_2)$ within the BHM.
 
As these frequencies can be measured with high accuracy, we are able to observe deviations from the simple BHM. We obtain a first correction by the inclusion of nearest-neighbor interactions \cite{Auerbachbook} in an extended two-site Bose-Hubbard model (EBHM, see Eq.~S1 in ref. \cite{SOMNote}). This modification introduces the inter-well interaction energy $U_{\rm LR} = g \times \int w^2_{\rm L}(\vec x)w^2_{\rm R}(\vec x)d^3 x$ and a correction to the tunneling matrix element, which becomes $J' = J+\Delta J$, where $\Delta J = -g \times \int w^3_{\rm L}(\vec x)w_{\rm R}(\vec x)d^3 x$. The inter-well interaction leads to a direct spin-exchange term, which in the limit $U \gg J$ reduces the corrected superexchange coupling to $J'_{\rm ex} = 2J'^2/U - U_{\rm LR}$. While we find that the corrections to the pure two-site Bose-Hubbard model are not negligible in the experimentally relevant parameter region, numerical calculations based on the multiband Schr\"{o}dinger equation show that the direct exchange can never overcome the superexchange coupling term (see Fig.~S1) and therefore change the nature of the ground state to be antiferromagnetic. This is in agreement with the Lieb-Mattis theorem \cite{Lieb:1962}, which states that the groundstate for two bosons has to be a spin-triplet state.

\paragraph*{Initial state preparation.}
In order to investigate the spin dynamics between neighboring atoms, we initially prepare a sample of ultracold neutral atoms with two relevant internal states $\spinup$ and $\spindown$ in a 3D array of double wells with N\'{e}el-type antiferromagnetic order $|\!\!\uparrow\downarrow\uparrow\downarrow\uparrow\downarrow\!\cdots\rangle$ along one spatial direction (see Fig.~2A). State preparation was started by loading a $^{87}{\rm Rb}$ Bose-Einstein condensate of typically around $8\times 10^{4}$ atoms in the $|F=1,m_F=-1\rangle$ Zeeman sublevel with no discernible thermal fraction from a magnetic trap with high offset field into a 3D optical lattice of double well potentials \cite{Sebby:2006}. This "superlattice" potential is obtained by superimposing on one axis two standing light fields with periodicity $382.5\,{\rm nm}$ (short lattice) and $765\,{\rm nm}$ (long lattice) and additional standing waves with periodicity $420\,{\rm nm}$ on the two perpendicular axes \cite{Foelling:2007a}. Controlling all depths and the relative phase of the short and long lattice allows to tune the double well configuration in terms of the Hamiltonian parameters $J$, $U$ and $\Delta$. The depths of the lattices are given in units of the short-lattice recoil energy $E_r=h^2/(2M\lambda^2)$ with $\lambda = 765\,{\rm nm}$ throughout the article. The loading ramps were optimized to favor an occupation of two atoms per double well and to avoid heating to higher vibrational levels \cite{SOMNote}. After merging the double wells by ramping down the short lattice, a microwave rapid adiabatic passage was used to transfer all atoms into the $|F=1,m_F=0\rangle$ state. Subsequently the magnetic trap is switched off and while keeping a homogeneous offset field of around $1.2\,{\rm G}$, atom pairs were coherently transferred from $|m_F=0;m_F=0\rangle$ into spin triplet pairs $(|{}{+1};-1\rangle+|{}{-1};+1\rangle)/\sqrt{2}$ by means of spin-changing collisions \cite{Widera:2005,Gerbier:2006c}. The two magnetic sublevels $|m_F=\pm 1\rangle$ correspond to the two spin states $\spinup$ and $\spindown$. The remaining atoms in the $|m_F=0\rangle$ state, e.g. on singly occupied sites, are transferred into $|F=2,m_F=0\rangle$ and removed in a filtering sequence before the detection \cite{SOMNote}. 

Finally, the short lattice was ramped up slowly in $20\,{\rm ms}$ thereby inhibiting a coherent splitting of the atom pairs and leaving the double wells in a state with one atom on each side \cite{Sebby:2007a}. For the time of the ramp up, a magnetic field gradient of $B' \approx 17\,{\rm G/cm}$ in the direction of the superlattice is switched on. Therefore, the degeneracy of the states $\spinupdown$ and $\spindownup$ in the double well is lifted by approximately $900\,{\rm Hz}$, which enables an adiabatic loading of the state $\spinupdown$ during the splitting process (see Fig.~2B). Numerical integration of a multiband ansatz for this procedure yields an expected fidelity of $> 99\%$ for creating an antiferromagnetic order along the axis of the superlattice. The mean population imbalance $x(t)$ and spin imbalance $N_z(t)$ of the ensemble of double wells was detected by applying a mapping technique \cite{Sebby:2007a,Foelling:2007a} combined with a Stern-Gerlach filter (see Fig.~2C). A maximum spin imbalance of 60-70\% was observed for our initial state corresponding to a probability of 80-85\% for having prepared the desired state $\spinupdown$. We believe that this measured value is mainly reduced due to our detection method. Direct spin-exchange processes emerge during the mapping sequence \cite{Anderlini:2007} and can lead to a mixing of the spin configuration and thus a reduction of the measured N\'{e}el order parameter \cite{SOMNote}.

\paragraph*{Time-resolved observation of superexchange interactions.}
The spin dynamics are initiated by rapidly ramping down the short lattice and thereby the double well barrier in $200\,{\rm \mu s}$, thus significantly increasing the tunneling and superexchange couplings. After letting the system evolve for a hold time $t$, the spin-configuration was frozen out by ramping up the barrier in $200\,{\rm \mu s}$, quenching both $J$ and $J_{\rm ex}$ again. The measurement of the ensemble averages $x(t)$ and $N_z(t)$ is carried out as described above.

Three typical time traces obtained by this procedure are shown in Figure~3. For low barrier depths ($J/U > 1$), we observe a pronounced time evolution of the spin imbalance $N_z(t)$ consisting of two frequency components with comparable amplitudes and frequencies (Fig.~3A). With increasing interaction energy $U$ relative to $J$, the frequency ratio increases, leaving a slow component with almost full amplitude and an additional high-frequency modulation with small amplitude (Fig.~3B). The fast component corresponds to first order tunneling due to the coupling of $|t\rangle$ and $|+\rangle$, which becomes more and more off-resonant as the barrier height is increased and therefore $J/U$ is decreased. For $J/U \ll 1$, it is completely suppressed and the only process visible is the superexchange oscillation (Fig.~3C). For all barrier heights, the population imbalance $x(t)$ stays flat, emphasizing that even though strong spin currents are present in the system, no net mass flow can be observed for our initial state. 
We fit the traces for $N_z(t)$ with a sum of two damped sine waves with variable frequencies $\omega_{1,2}/(2\pi)$ and amplitudes $A_{1,2}$. For the damping we assume a Gaussian characteristics with $1/e$-damping constants $\gamma_{1,2}$. The results of the fit are displayed in Fig.~4. For $V_{\rm short} \geq 15\,E_r$ (inset in Fig.~4A) we can identify only a single frequency component corresponding to the superexchange oscillation with $4J^2/(h U)$ and full relative amplitude (see Fig.~4B). We are able to observe this frequency down to $4.8(4)\,{\rm Hz}$ at $J/U = 0.023$ for $V_{\rm short}=20\,E_r$. The damping of the signal can be explained by the inhomogeneous distribution of coupling parameters due to the Gaussian shape of the lattice beams, which leads to a dephasing of the evolution within the ensemble. For $V_{\rm short} \geq 17E_r$, additional damping mechanisms like tunneling to empty adjacent lattice sites (defects) or small residual inhomogeneous magnetic field gradients become relevant and limit the measurements (see Fig.~4C).

The comparison of the results with the theoretical predictions by the simple BHM shows significant deviations at low barrier heights which cannot be explained by our uncertainties in the lattice depths. In this region, the EBHM can model the experimental data much more accurately. This can be understood by the fact that the inter-well interaction energy increases with decreasing barrier and begins to noticeably influence the dynamics \cite{SOMNote}. In fact, the EBHM description yields $\hbar(\omega_1-\omega_2) = U+3U_{\rm LR}$ and therefore directly explains the upward trend of this frequency difference for small short-lattice depths (see Fig.~S2). For large barrier heights our experimental data is compatible with both models within the uncertainties of the lattice depths. However, here, the predictions of the EBHM are always closer to the measured values.

\paragraph*{Sign reversal of the effective coupling parameter.}
In order to demonstrate the controllability of superexchange interactions, the spin dynamics was investigated with an applied bias on the double wells for a short lattice depth of $15\,E_r$ and the same depths for the long and transverse lattice as before. Starting with an initial antiferromagnetically ordered state, as above, we first let the system evolve in symmetric double wells ($\Delta=0$) for $t_0=4.5\,{\rm ms}$ until the first node $N_z(t)=0$ of the spin imbalance is reached for the state $(|t\rangle + i|s\rangle)/\sqrt{2}$ (see Fig.~5). After freezing out the relative phase between $|s\rangle$ and $|t\rangle$ by ramping up the potential barrier, a defined potential bias $\Delta$ is applied and a second evolution sequence with hold time $t'=t-t_0$ is initialized by ramping down the short lattice again to $15\,E_r$. The subsequent detection follows the scheme described above.

Fig.~5A shows the evolution of the spin imbalance $N_z(t)$ in symmetric double wells together with the time traces for two different bias energies $\Delta >U$ yielding an effective coupling of $J_{\rm ex}(\Delta) \approx -J_{\rm ex}(\Delta=0)$ and $-J_{\rm ex}(\Delta=0)/2$ respectively. The sign reversal of $J_{\rm ex}$ due to the introduction of the bias is directly visible by the change in slope of the spin imbalance $t=t_0$. It should be noted that the now negative sign of $J_{\rm ex}$ for bosons does not imply a violation of the Lieb-Mattis theorem, as the new ground state in this regime is the spin triplet state $\spinpairleft$ and the superexchange couples the first and second excited states $|s\rangle$ and $|t\rangle$ which have reversed order for $\Delta > U$ (see left inset in Fig.~5B).

The introduction of a non-zero tilt leads to an increased sensitivity of the exchange frequency to fluctuations due to the inhomogeneities in the array of double wells which are most effective around $\Delta \approx U$. Therefore the damping of the signal due to dephasing is stronger as $\Delta$ approaches $U$. A fit of a single dampened sine wave to the time traces obtained for various tilts yields the frequency curve shown in Fig.~5B together with the amplitude of the oscillation. Starting around $50\,{\rm Hz}$, the oscillation frequency reaches a resonance for $\Delta / U \approx 1$, where the amplitude reverses sign, leading to the observed time reversal in the dynamics.

\paragraph*{Summary and outlook.}         
We have demonstrated time-resolved measurements of  superexchange spin interactions between ultracold atoms on neighboring lattice sites and have shown how to control such interactions with optical superlattices. Comparing the measurements to theoretical predictions of these spin interactions from first principles we find excellent agreement of our data to an extended two-site version of the Bose-Hubbard model. Although superexchange interactions become exponentially suppressed for deep optical lattices, the coupling strength $2J_{\rm ex}/h$ can be several hundred Hertz for lattice depths of around $12-15\,E_r$ and thus almost a factor of thousand larger than the direct magnetic dipole-dipole interaction of Rb atoms on neighboring lattice sites. One order of magnitude larger coupling strengths than the ones shown here, however, could still be achieved using electric dipole-dipole mediated spin interactions between ground state polar molecules \cite{Micheli:2006}.

The demonstrated scheme to change the superexchange coupling strength and reverse the sign of the spin interaction can also be applied to the full 1D chain, offering novel possibilities for engineering spin-spin interactions in optical lattices. It is now e.g. conceivable to engineer a setup with ferromagnetic interactions along one and antiferromagnetic interactions along another lattice direction. Furthermore, one can dynamically switch between ferro- or antiferromagnetic interactions along a given lattice direction and follow the subsequent dynamical evolution of the quantum spin system. 

When the presented loading scheme is carried out without any magnetic gradient field during the splitting process, a valence-bond solid (VBS) type spin state \cite{Auerbachbook,Alet:2006} can be efficiently engineered. Such VBS states can be viewed as a large array of robust Bell pairs \cite{Roos:2004,Langer:2006}. In principle, the superexchange interaction can be changed to be of Ising-type, e.g. by tuning the interspecies scattering length \cite{Duan:2003}. Thereby, it can be used to create large entangled states out of the initially disconnected pairs, which have been shown to be powerful resources for measurement based quantum computation \cite{Briegel:2001,Verstraete:2004}. Moreover, controlling the superexchange interactions along different lattice directions also offers novel possibilities for the generation of topological many-body states for quantum information processing \cite{Kitaev:2006,Duan:2003}.

\begin{figure}[p]
\begin{center}
\includegraphics[width=0.9\linewidth]{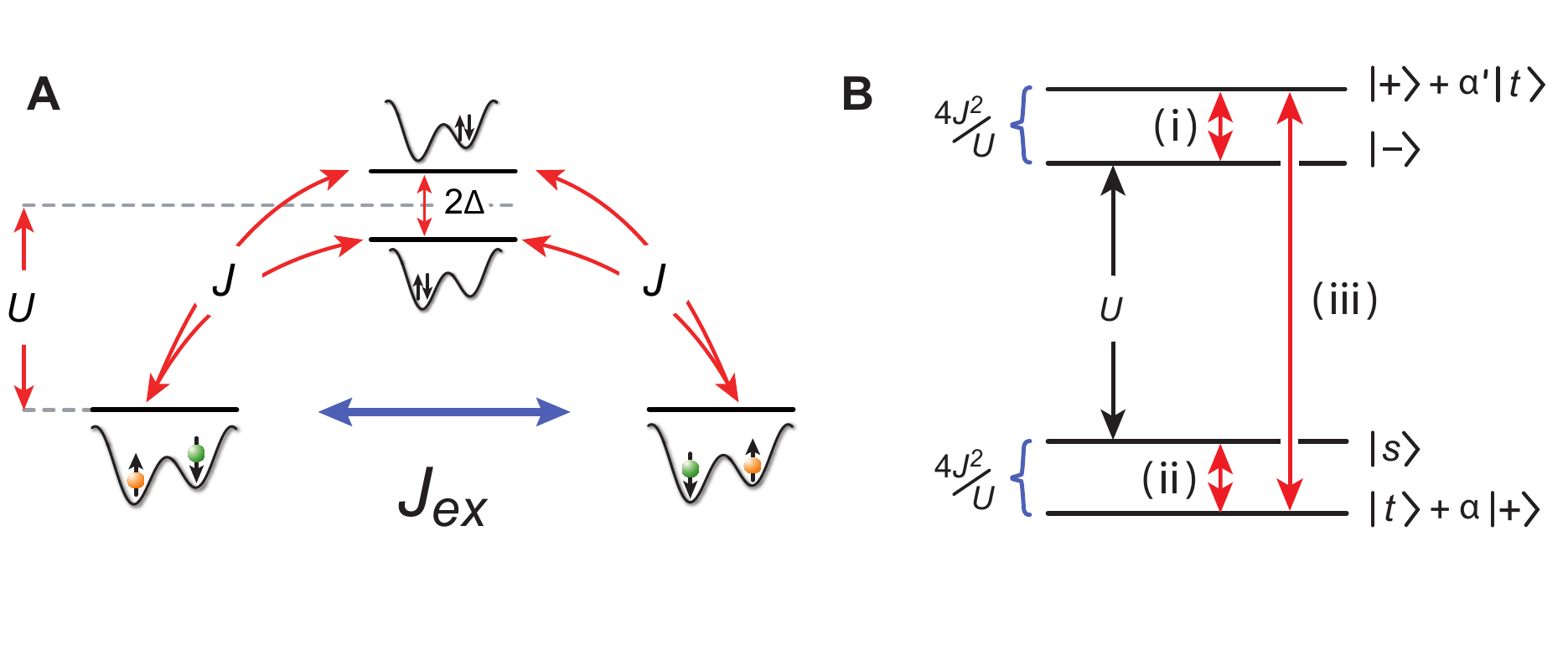}
\caption{Schematics of superexchange interactions. {\bf A} Second order hopping via $\spinpairleft$ and $\spinpairright$ mediates the spin-spin interactions between atoms on different sides of the double well. {\bf B} Energy levels for $\Delta=0$ and $U \gg J$. The evolution in the upper doublet of states (i) corresponds to the correlated tunneling of atom pairs \cite{Foelling:2007a}, while the superexchange takes place in the lower one (ii). Both doublets are coupled by first order tunneling processes (iii).}
\end{center}
\end{figure}

\begin{figure}[p]
\begin{center}
\includegraphics[width=0.9\linewidth]{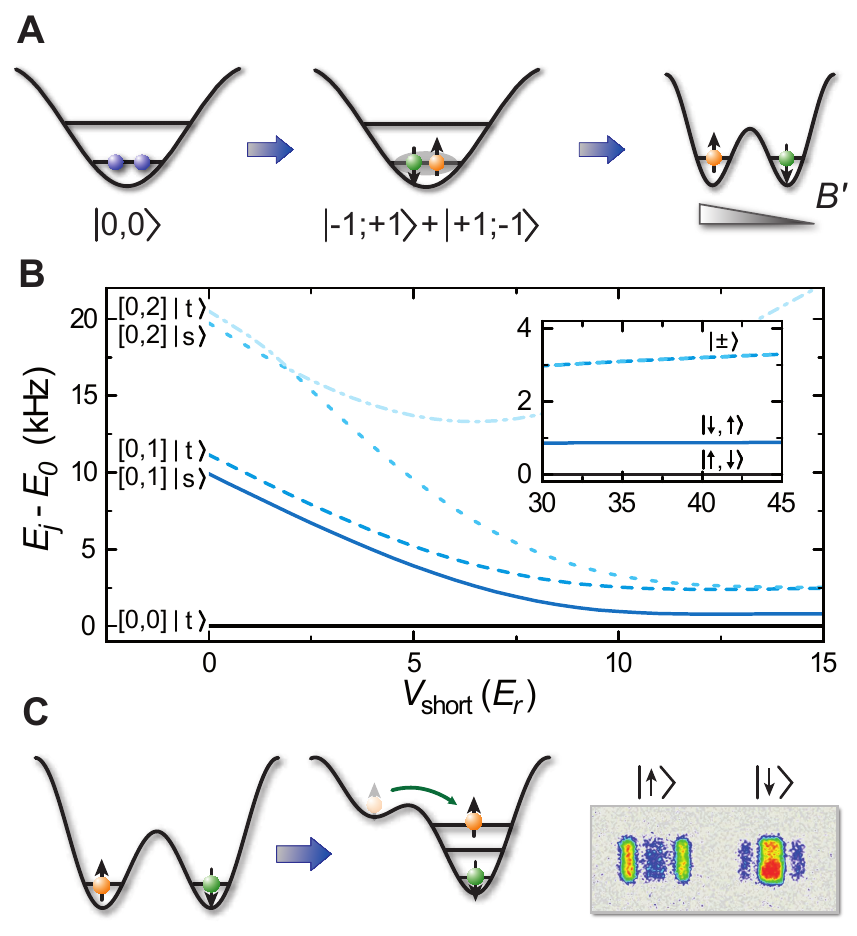}
\caption{State preparation and detection. {\bf A} Spin triplet pairs are created on doubly occupied lattice sites and subsequently split under the influence of a magnetic field gradient $B'$ to obtain antiferromagnetic N\'{e}el order. {\bf B} Evolution of the eigenenergies with respect to the ground state energy during the splitting with $V_{\rm long} = 10\,E_r$, $V_{\rm trans} = 25\,E_r$ and a gradient of $B'=17{\rm G/cm}$. The notation $[v_1,v_2]$ denotes the number of vibrational excitations for the first and second particle. {\bf C} Detection of the population and spin imbalance. The population of the left well is transferred to a higher vibrational level of the underlying long-lattice well \cite{Sebby:2006,Foelling:2007a}. Subsequent band mapping and a Stern-Gerlach filter allow to determine $x(t)$ and $N_z(t)$ from the time-of-flight (TOF) images.}
\end{center}
\end{figure}

\begin{figure}[p]
\begin{center}
\includegraphics[width=0.9\linewidth]{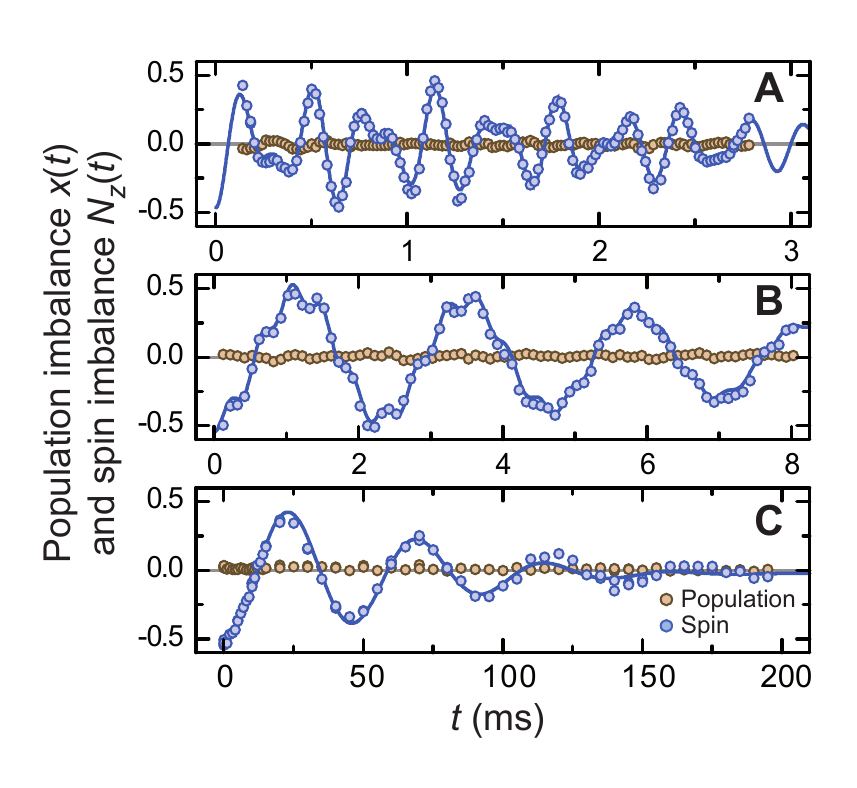}
\caption{Spin and population dynamics in symmetric double wells. The time evolution of the mean spin $N_z(t)$ (blue circles) and population imbalance $x(t)$ (brown circles) are shown for three barrier depths within the double well potential: ({\bf A}) $V_{\rm short}=6\,E_r, J/U=1.25$, ({\bf B}) $V_{\rm short}=11\,E_r, J/U=0.26$ and ({\bf C}) $V_{\rm short}=17\,E_r, J/U=0.048$. The measured traces for the spin imbalance are fitted with the sum of two damped sine waves (blue lines). The population imbalance $x(t)$ stays flat for all traces.}
\end{center}
\end{figure}

\begin{figure}[p]
\begin{center}
\includegraphics[width=0.9\linewidth]{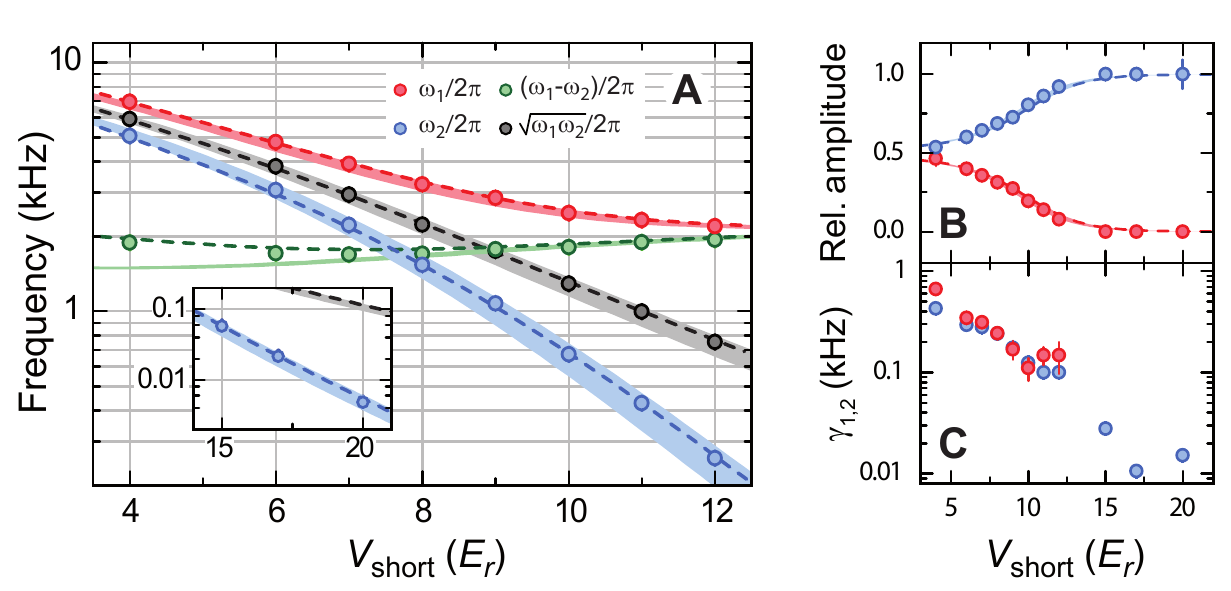}
\caption{Frequency components of the spin oscillations. {\bf A} The frequencies obtained by fitting the spin imbalance data for various values of $V_{\rm short}$ (red and blue circles) as well as the frequency difference and the geometrical mean (green and black circles) are compared to the theoretical predictions of the simple BHM (colored regions). The width of the regions represents a 2\% uncertainty in all lattice depths. The $2\sigma$ error bars on the fit results and extracted values are in all cases smaller than the data points. The dashed lines are the predictions of the EBHM. {\bf B} The plot of relative amplitudes vs. short lattice depth shows the suppression of the first order tunneling process towards small values of $J/U$. The damping coefficients for the two oscillatory components vs. lattice depth are displayed in {\bf C}.}
\end{center}
\end{figure}

\begin{figure}
\begin{center}
\includegraphics[width=0.9\linewidth]{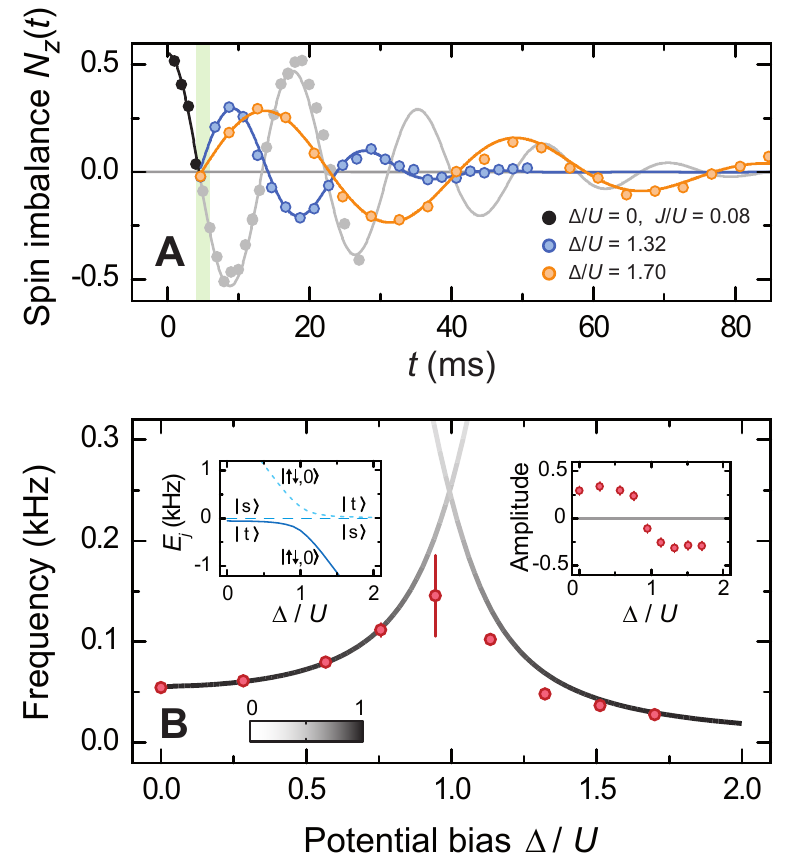}
\caption{Sign reversal of the superexchange coupling. {\bf A} After a quarter of a superexchange oscillation in symmetric double wells (black and gray dots, $J/U = 0.08$), a defined bias is applied, leading to a reversal in the slope of the spin imbalance and ongoing oscillations. For $\Delta / U = 1.32$ (blue circles), the magnitude of the coupling is almost the same as for $\Delta = 0$, whereas it is decreased to $\sim 50\%$ for $\Delta/U=1.70$ (orange circles). {\bf B} The fit of the data for various tilts to a damped sine yields a resonance in frequency at $\Delta \approx U$, where the amplitude reverses sign (see right inset). The solid lines show the theoretically expected frequencies for the given parameters where the shading reflects the amplitude of the corresponding Fourier component. The left inset shows the three lowest eigenenergies of the Bose-Hubbard Hamiltonian versus the potential bias.}
\end{center}
\end{figure}

\pagebreak

\bibliographystyle{Science}

\begin{thebibliography}{10}
\bibitem{ScattLengthNote}
 For the spin-states used in our experiment, the interaction energies for the
  different intercombinations of spins vary by only a few percent.
\bibitem{SOMNote}
 Further materials and methods are available as supporting material.
\end{thebibliography}

\begin{scilastnote}
\item We acknowledge helpful discussions with B. Paredes and funding by the DFG, the EU (OLAQUI, SCALA), NSF, AFOSR, MURI and the Packard Foundation. S.T. aknowledges financial support by MATCOR.
\end{scilastnote}
\clearpage

% SOMSOMSOMSOMSOMSOMSOMSOMSOMSOM
% SOMSOMSOMSOMSOMSOMSOMSOMSOMSOM
% SOMSOMSOMSOMSOMSOMSOMSOMSOMSOM
% SOMSOMSOMSOMSOMSOMSOMSOMSOMSOM
% SOMSOMSOMSOMSOMSOMSOMSOMSOMSOM

\setcounter{figure}{0}
\setcounter{equation}{0}
\renewcommand{\theequation}{S\arabic{equation}}
\renewcommand{\thefigure}{S\arabic{figure}}

\begin{center}
{\LARGE Supporting Online Material for:\\[0.5cm]Time-resolved Observation and Control of\\ Superexchange Interactions with Ultracold Atoms in\\ Optical Lattices}\\[1cm]
{\large S. Trotzky,$^{1\dagger}$ P. Cheinet,$^{1\dagger}$ S. F\"olling,$^{1}$ M. Feld,$^{1,2}$\\ U. Schnorrberger,$^{1}$ A. M. Rey,$^{3}$ A. Polkovnikov,$^{4}$ E. A. Demler,$^{3,5}$\\ M. D. Lukin,$^{3,5}$ and I. Bloch$^{1\ast}$}\\[0.8cm]
\normalsize{$^{1}$Institut f\"ur Physik, Johannes Gutenberg-Universit\"{a}t, 55099 Mainz, Germany}\\
\normalsize{$^{2}$Fachbereich Physik, Technische Universit\"{a}t Kaiserslautern, 67663 Kaiserslautern, Germany}\\
\normalsize{$^{3}$Institute for Theoretical Atomic, Molecular and Optical Physics,}\\
\normalsize{Harvard-Smithsonian Center of Astrophysics, Cambridge, MA, 02138, USA}\\
\normalsize{$^{4}$Department of Physics, Boston University, Boston, MA, 02215, USA}\\
\normalsize{$^{5}$Physics Department, Harvard University, Cambridge, MA, 02138, USA}\\[0.8cm]
\normalsize{$^\ast$To whom correspondence should be addressed; E-mail:  bloch@uni-mainz.de.}\\
\normalsize{$^\dagger$These authors contributed equally to this work.}
\end{center}

\paragraph*{Extended two-site Bose-Hubbard model.}
The inclusion of next-neighbor interactions to the simple BHM Hamiltonian defined in Eq.~S1 leads to the extended two-site Bose-Hubbard Hamiltonian (\emph{S1})
\begin{eqnarray}
    \hat H^{\rm EBHM} 
 &=&\hat H^{\rm BHM} 
  - \Delta J \sum_{\sigma \neq \sigma'} 
          \left(\hat n_{\rm \sigma L}+\hat n_{\rm \sigma R}\right)
          \left(\hat a_{\rm \sigma' L}^\dagger \hat a_{\rm \sigma' R}
               +\hat a_{\rm \sigma' R}^\dagger \hat a_{\rm \sigma' L}\right)\nonumber\\
 &+&U_{\rm LR} \sum_{\sigma \neq \sigma'}
          \left(\hat n_{\rm \sigma L}\hat n_{\rm \sigma' R}
               +\hat a_{\rm \sigma L}^\dagger \hat a_{\rm \sigma'R}^\dagger
                \hat a_{\rm \sigma'L} \hat a_{\rm \sigma R}\phantom{\frac{1}{2}}\right.\nonumber\\
 &&\phantom{U_{\rm LR} \sum_{\sigma \neq \sigma'}}\left.
               +\frac{1}{2}\hat a_{\rm \sigma L}^\dagger \hat a_{\rm \sigma'L}^\dagger
                \hat a_{\rm \sigma'R} \hat a_{\rm \sigma R}
               +\frac{1}{2}\hat a_{\rm \sigma R}^\dagger \hat a_{\rm \sigma'R}^\dagger
                \hat a_{\rm \sigma'L} \hat a_{\rm \sigma L}\right)\,,
\end{eqnarray}
where the parameters $\Delta J$ and $U_{\rm LR}$ are defined in the main text. Here, we focus on a system of two coupled wells occupied by exactly two atoms in the two different spin-states $\spinup$ and $\spindown$. The relation of $\Delta J$ and $U_{\rm LR}$ to the tunneling matrix element $J$ is plotted in Fig.~S1. It is apparent that those terms in Eq.~S1 proportional to $\Delta J$ act in the same way on an arbitrary Fock state in the system as the tunneling operator does and therefore modify $J$ to now $J' = J+\Delta J$, while the terms proportional to $U_{\rm LR}$ lead to an energy shift of the states with exactly one atom on each site with respect to the states with double occupancy in a single well. The states $|s\rangle$ and $|-\rangle$ stay eigenstates of the system and the states $|t\rangle$ and $|+\rangle$ are now coupled via the matrix
\begin{equation}
  H^{\rm EBHM}_{t,+} =
  \left(\begin{array}{cc}2U_{\rm LR} & -2J^\prime \\
                         -2J^\prime  & U+U_{\rm LR}\end{array}\right)\,
\end{equation}
with the eigenvalues
\begin{equation}
  \pm\hbar\omega_{1,2}^{\rm EBHM} = \frac{U-U_{\rm LR}}{2}\left(1\pm\sqrt{\left(\frac{4J'}{U-U_{\rm LR}}\right)^2+1}\right) + 2U_{\rm LR}\,,
\end{equation}
which correspond to the Fourier components for the evolution of the initial state $\spinupdown$ under the Hamiltonian Eq.~S1. In the regime of strong interactions, the frequency $\omega_2$ is assigned to the superexchange process and by means of perturbation theory up to second order one finds the effective coupling parameter to be $J'_{\rm ex} = 2J'^2/U - U_{\rm LR}$. The direct spin-exchange term $U_{\rm LR}$ favors an antiferromagnetic ground state for repulsively interacting bosons. However, the inset in Fig.~S1 shows, that the case $U_{\rm LR} > 2J'^2/U$ is never reached and therefore the presence of the direct nearest neighbor interactions does not lead to a violation of the Lieb-Mattis
theorem (\emph{S2}). With $V_{\rm long} = 10 E_r$ and $V_{\rm trans} = 41 E_r$, the dependency of the two contributions on the short-lattice depth is well approximated by two exponential laws of the type $B_{\rm s,d} \times \exp(-\frac{1}{2} V_{\rm short} / E_r)$, where $B_{\rm s} = 15.24 E_r$ for the superexchange term and $B_{\rm d} = 0.36 E_r$ for the direct exchange term. 

\paragraph*{Loading sequence.}
Since the short lattice is created by a Gaussian laser beam at a wavelength of $765\,{\rm nm}$, it is blue-detuned with respect to the rubidium D1 and D2 transition and thus creates an anti-confining potential in transversal direction. The simultaneous ramp up of the short lattice ($\sim 25\,E_r$) together with the red-detuned transversal lattices ($840\,{\rm nm}$ laser wavelength, $\sim 41\,E_r$) leads to a spreading of the atom distribution before freezing out tunneling and therefore to a more homogeneous filling than would be obtained by a fully red-detuned 3D optical lattice with the same beam waist and atom number. Subsequently, the long lattice is ramped up to $\sim 10\,E_r$ to obtain isolated double wells. The lattice depths during the loading and the shape of the ramps have been optimized to give mean occupation of each double well sites close to two. By means of an interferometric measurement similar to the one described in (\emph{S3}), we verify that about 70-80\% of the atoms are loaded to doubly occupied sites which is compatible with the conversion efficiency we observe in the spin-changing collisions.

\paragraph*{Detection and filtering sequence.} 
For each double well, we transfer the population of the left well into the second excited band of the underlying long lattice well by applying a tilt and ramping down the barrier afterwards (\emph{S3,S4}). A subsequent adiabatic band mapping (\emph{S5,S6}) results in spatial separation of the atoms initially localized on either side of the double well after TOF. A Stern-Gerlach filter at the beginning of TOF allows for a spin-state selective detection. In addition, we transfer those atoms that have not undergone the spin-changing collisions and therefore stayed in $|F=1,m_F=0\rangle$ with a microwave $\pi$-pulse to $|F=2,m_F=0\rangle$. They are removed by a resonant laser pulse during time-of-flight expansion to obtain the signal only from atoms in doubly occupied sites.

Since the wave functions of both atoms are brought to overlap during the detection sequence, a direct exchange term emerges which leads to spin-oscillations between the atoms initially located on separate sites (\emph{S7}). To minimize this effect, we carry out the band mapping immediately ($\simeq 70\,{\rm \mu s}$) after the transfer. Since we cannot suppress the direct spin-exchange oscillations completely in our sequence, we believe these to limit our detection efficiency, while the fidelity of the preparation is probably higher and therefore closer to the expected 99\%. 

\paragraph*{Validity of the extended two-site Bose-Hubbard model.}
We investigate the validity of the EBHM by means of the frequency difference $\omega_1 - \omega_2$ in Fig.~S2. While for low barrier depth, the EBHM models the data much better than the simple BHM, we still find deviations from its predictions which exceed the 2\% uncertainty in our lattice depth. One reason for this can be the fact that the parameters for the BHM as well as for the EBHM are obtained by a single particle band structure calculation which does not account for the atom-atom interaction in the pairs. For a repulsive interaction and atom numbers larger than $1$, a broadening of the single particle wave function has to be taken into account, which leads to lower onsite and higher intersite interaction energies (\emph{S8,S9}). The comparison of the measurements to the solution of the two-band Schr\"{o}dinger equation which includes the interaction from the start already shows a much better agreement. When more bands are taken into account, the two-frequency approximation used to fit the data is not valid anymore a direct comparison to the fit results is therefore not reasonable. 

\begin{figure}[p]
\begin{center}
\includegraphics[width=0.9\linewidth]{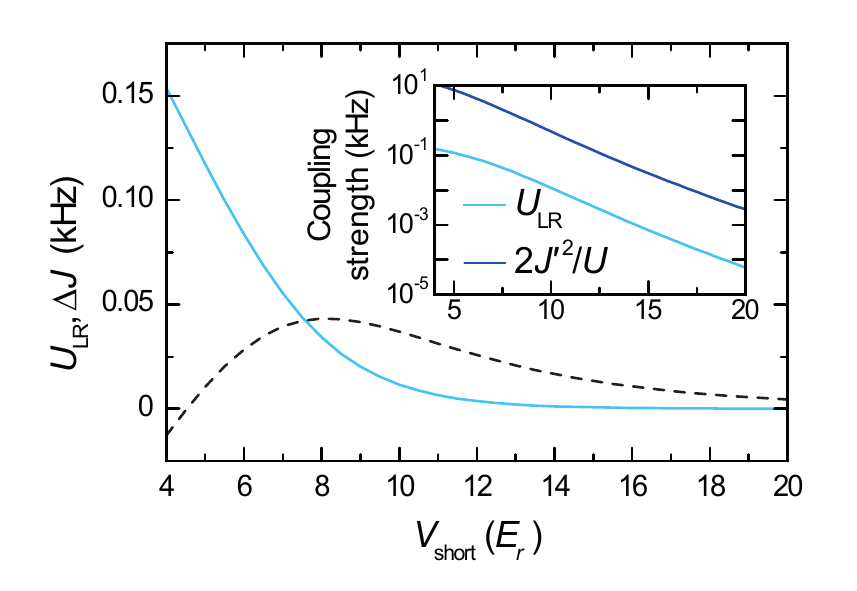}
\caption{Extended Bose-Hubbard parameters. The parameters $U_{\rm LR}$ (light blue curve) and $\Delta J$ (black dashed curve) are plotted versus $V_{\rm short}$. The inset shows a comparison of the direct exchange term $U_{\rm LR}$ (light blue) to the 
superexchange term $2 J'^2/U$ (dark blue), where the latter is always at least one order of magnitude larger than the first.}
\end{center}
\end{figure}

\begin{figure}[p]
\begin{center}
\includegraphics[width=0.9\linewidth]{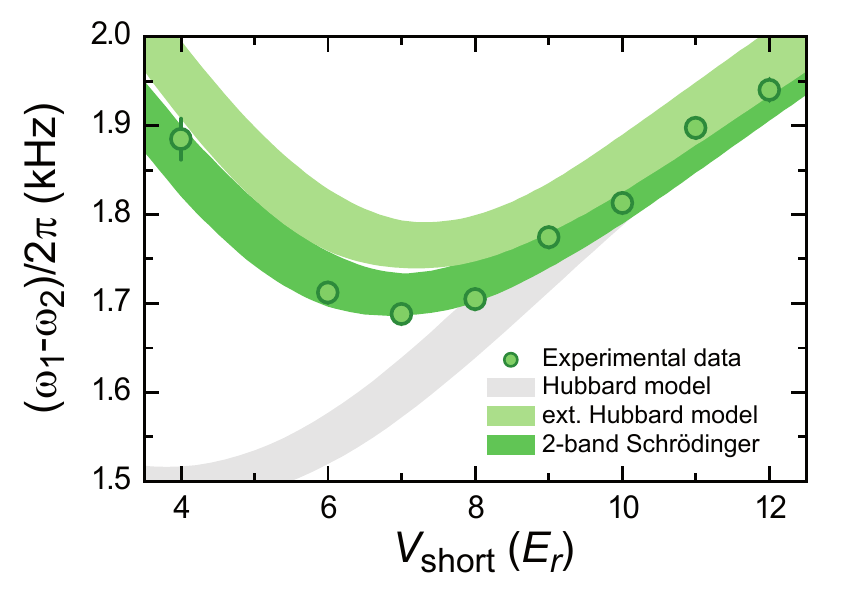}
\caption{Comparison of the different models. The difference of the fitted frequencies $\omega_{1,2}/(2\pi)$ versus $V_{\rm short}$ are plotted as in Fig.~4A and compared to the predictions of the simple BHM (grey region, $U/h$), the EBHM (light green region, $(U+3U_{\rm LR})/h$) and the solution of the two-band Schr\"{o}dinger equation (dark green region). The width of the regions reflects the 2\% uncertainty in lattice depths and the error bars on the data points denote the 95\% confidence interval as obtained from the fits.}
\end{center}
\end{figure}

\clearpage

{\noindent\bf\Large References and Notes}
\begin{enumerate}
\item[S1.]
V.~Scarola, S.~D. Sarma, {\it Phys. Rev. Lett.\/} {\bf 95}, 033003 (2005).
\item[S2.]
E.~Lieb, D.~Mattis, {\it Phys. Rev.\/} {\bf 125}, 164 (1962).
\item[S3.]
J.~Sebby-Strabley, {\it et~al.\/}, {\it Phys. Rev. Lett.\/} {\bf 98}, 200405
  (2007).
\item[S4.]
S.~F{\"o}lling, {\it et~al.\/}, {\it Nature\/} {\bf 448}, 1029 (2007).
\item[S5.]
A.~Kastberg, W.~D. Phillips, S.~L. Rolston, R.~J.~C. Spreeuw, P.~S. Jessen,
  {\it Phys. Rev. Lett.\/} {\bf 74}, 1542 (1995).
\item[S6.]
M.~Greiner, I.~Bloch, M.~O. Mandel, T.~H{\"a}nsch, T.~Esslinger, {\it Phys.
  Rev. Lett.\/} {\bf 87}, 160405 (2001).
\item[S7.]
M.~Anderlini, {\it et~al.\/}, {\it Nature\/} {\bf 448}, 452 (2007).
\item[S8.]
G.~Campbell, {\it et~al.\/}, {\it Science\/} {\bf 313}, 5787 (2006).
\item[S9.]
J.~Li, Y.~Yu, A.~M. Dudarev, Q.~Niu, {\it New J. Phys.\/} {\bf 8}, 154 (2006).
\end{enumerate}

\end{document}